\documentclass[12pt]{article}
\usepackage{epsfig,cite,amssymb,graphicx}

\makeatletter
\@addtoreset{equation}{section}

\makeatletter
\renewcommand\section{\@startsection {section}{1}{\z@}%
                                   {-3.5ex \@plus -1ex \@minus -.2ex}
                                   {2.3ex \@plus.2ex}%
                                   {\normalfont\large\bfseries}}
\renewcommand\subsection{\@startsection{subsection}{2}{\z@}%
                                     {-3.25ex\@plus -1ex \@minus -.2ex}%
                                     {1.5ex \@plus .2ex}%
                                     {\normalfont\bfseries}}

\def\baselinestretch{1.2}
\newcommand{\be}{\begin{equation}}
\newcommand{\ee}{\end{equation}}
\newcommand{\beq}{\begin{eqnarray}}
\newcommand{\eeq}{\end{eqnarray}}
\newcommand{\R}{{\mathbb R}}
\newcommand{\Z}{{\mathbb Z}}
\newcommand{\U}{{\cal U}}
\newcommand{\M}{{\cal M}}
\newcommand{\W}{{\cal W}}

\newcommand{\FIGURE}[2][v]{\begin{figure}[#1]#2\end{figure}}

\begin{document}
\begin{titlepage}

\begin{flushright}
hep-th/0210105\\
\end{flushright}

\vfil\vfil

\begin{center}

{\large{{\bf  Supergravity Solution of Intersecting Branes \\ and AdS/CFT with Flavor}}
}

\vfil

\vspace{5mm}

Sergey A. Cherkis and Akikazu Hashimoto\\

\vspace{10mm}

Institute for  Advanced Study\\ School of Natural Sciences\\
Einstein Drive, Princeton, NJ 08540\\

\vfil

\end{center}

\begin{abstract}
\noindent We construct the supergravity solution for fully localized
D2/D6 intersection.  The near horizon limit of this solution is the
supergravity dual of supersymmetric Yang-Mills theory in 2+1
dimensions with flavor.  We use this solution to formulate mirror
symmetry of 2+1 dimensional gauge theories in the language of AdS/CFT
correspondence.  We also construct the supergravity dual of a
non-commutative gauge theory with fundamental matter.
\end{abstract}

\vspace{0.5in}

\end{titlepage}
\renewcommand{\baselinestretch}{1.05}  

\section{Introduction}

Supergravity solutions of intersecting D-branes are relatively easy to
find as long as they are sufficiently smeared \cite{Tseytlin:1996bh}.
Supergravity solutions of the localized intersections are far more
difficult to find. Starting with the work of \cite{Tseytlin:1997cs}
there has been a steady enterprise of attempts to construct such
supergravity solutions 
\cite{
Gauntlett:1997pk,
Gauntlett:1997cv,
Itzhaki:1998uz,
Hashimoto:1998ug,
Cherkis:1999jt,
Surya:1998dx,
Marolf:1999uq,
Gomberoff:1999ps,
Loewy:1999mn,
Yang:1999ze,
Hosomichi:2000iz,
Kaya:2000zs,
Rajaraman:2000ws,
Fayyazuddin:1999zu,
Brinne:2000fh,
Fayyazuddin:2002bm,
Kehagias:1998gn,
Grana:2001xn
}. 
However, to date, there are no known techniques for determining the
gravitational back reaction due to a general intersecting brane
configurations that arise in string theory.

For brane intersections involving a D6-brane, there are special
techniques which allow the explicit construction of certain localized
intersections \cite{
Itzhaki:1998uz,
Hashimoto:1998ug,
Cherkis:1999jt
}.
Here, one takes advantage of the fact that the M-theory lift of the
geometry of the D6-brane near its core is an ALE space which is
essentially the flat $\R^4$.  This allowed the construction of the completely
localized supergravity solution of D2 parallel to D6, as well as D4
ending on the D6.  This method however was limited in its
applicability to the region near the core of the D6.

The aim of this paper is to construct the fully localized supergravity
solution of D2 parallel to D6 without restricting to the near core
region of the D6. The construction of the supergravity solution turns
out to be possible for this case due to the fact that there exists a
simple ansatz which reduces the problem to a single linear
differential equation \cite{
Tseytlin:1997cs, 
Gauntlett:1997pk }
which is separable \cite{Surya:1998dx} and admits a regular boundary
condition \cite{Marolf:1999uq}. Therefore, the problem can be solved
using elementary methods. The solutions we obtain are nonetheless very
interesting. Instead of taking the near horizon limit of the D6-brane
as was done in \cite{Itzhaki:1998uz,Marolf:1999uq}, one can consider
the near horizon limit of the D2-brane.  This gives rise to a geometry
where a D6-brane is slicing through the near horizon geometry of the
D2-branes. From the open string point of view, this corresponds to
taking the decoupling limit which keeps only the gauge fields on the
D2-brane and the charged fundamental matter arising from strings
stretching between the D2 and the D6 branes. It is therefore the 2+1
dimensional version of the holographic dual of gauge theory with
flavors considered recently by Katz and Karch in \cite{Karch:2002sh}
for the 3+1 dimensional case.  There are several advantages for
considering the case of 2+1 dimensions over 3+1.  The field theory in
2+1 dimensions is superrenormalizable even after adding the
fundamental matter, in contrast to 3+1 dimensional theory which looses
the asymptotic freedom.\footnote{There are, however, constructions
involving orientifolds which maintain conformal invariance
\cite{Fayyazuddin:1998fb,Aharony:1998xz}.}  The supergravity solution
takes the full gravitational back reaction of the D6-branes into
account in contrast to the 3+1 example where the effect of D7-brane is
treated in the probe approximation. One can therefore think of the
localized D2/D6 solution as the ``cleaner'' version of the
AdS/CFT-like correspondence with fundamental flavors.

The near horizon limit of the D2/D6 solution captures the full RG flow
of the weakly coupled supersymmetric Yang-Mills theory with
fundamental matter in the UV to a superconformal fixed point in the IR
along the lines of \cite{Itzhaki:1998dd}. In fact, certain qualitative
features of precisely this RG flow was anticipated in
\cite{Pelc:1999ms}. Our explicit supergravity solution confirms the
expectation of \cite{Pelc:1999ms}. As an added bonus, these
supergravity solutions can be used to illustrate the mirror symmetry
of Intriligator and Seiberg \cite{Intriligator:1996ex} in the language
of AdS/CFT correspondence.
 
\section{The Solution}

Let us begin by describing the explicit construction of the
supergravity solution. The idea is to start with a lift of D6-brane to
the Taub-NUT geometry in M-theory and to consider the effect of placing
large number of M2-branes in this background. To describe this background,
one employs the ansatz
\beq
ds^2 &=& H^{-2/3} (-dt^2 + dx_1^2 + dx_2^2) + H^{1/3} (dy^2 + y^2 d \Omega_3^2 + ds_{TN}^2) \nonumber \\
F &=& dt \wedge dx_1 \wedge dx_2 \wedge d H^{-1} \label{ansatz}
\eeq
where $ds_{TN}^2$ is the metric of the Taub-NUT space. We find it
convenient to parameterize the coordinates of the Taub-NUT space so
that the metric takes the form
\be ds_{TN}^2 = \left(1 + {2m \over r} \right) (d r^2 + r^2 (d \theta^2 + \sin^2\theta d \phi^2)) + \left({1 \over 1 + {2m \over r}}\right) (4m)^2
\left( d \psi + {1 \over 2} \cos \theta d \phi \right)^2 \label{TNmetric}\ee
where the coordinate take on values with range $0 \le r$, $0 \le
\theta \le \pi$, $0 \le \phi \le 2 \pi$, and $0 \le \psi \le 2 \pi$.
The parameter $m$ is related to the radius $R$ of the circle of the
Taub-NUT metric at infinity by the formula
\be R = 4 m \ .  \ee
For small radius 
\be r = {z^2  \over 8 m} \ll m \ , \ee
the Taub-NUT metric simplifies to
\be ds_{TN}^2 = dz^2 + z^2 d \Omega_3^2\ .  \ee
The ansatz (\ref{ansatz}) is a solution to the equation of motion of
11 dimensional supergravity if $H$ solves the harmonic equation in the
background of $\R^4 \times \mbox{Taub-NUT}$ space
\be \nabla^2 H = 0, \label{harm} \ee
except at the location of the M2-brane source. In order to maximize
the symmetry of the problem to simplify the analysis, let us consider
the case where the M2-brane source is placed at the origin $y = r =
0$. One can than take $H(y,r)$ to be a function of two variables, and
the harmonic equation (\ref{harm}) becomes
\be \left({1 \over 1+ {2 m \over r}}\right) \left({\partial^2 \over
\partial r^2} + {2 \over r}{\partial \over \partial r} \right)
H(y,r) + \nabla_y^2 H(y,r) =0\ . \ee
Our task is simply to solve this differential equation. To this end,
it is convenient to separate variables
\be H(y,r) = 1 + Q_{M2} \int {d^4 p \over (2 \pi)^4} e^{i p y} H_p(r)\ ,  \label{fourier} \ee
where $Q_{M2}$ is the membrane charge in the standard normalization 
\be Q_{M2} = 32 \pi^2 N_2 l_p^6 \ . \ee
Then,  $H_p$ satisfies
\be \left({1 \over 1+ {2 m \over r}}\right) \left({\partial^2 \over
\partial r^2} + {2 \over r}{\partial \over \partial r} \right)
H_p(r) -p^2  H_p(r) =0\ . \ee
Since this is a second order differential equation, there are two
solutions.  The one which decays at large $r$ can be written in a
closed analytic form
\be H_p(r) = c_p e^{-pr} \U(1+p m, 2, 2pr) \ , \label{analyticform}  \ee
where $\U(a,b,z)$ is the confluent hypergeometric function \cite{hmf}.
The normalization factor $c_p$ is fixed to
\be c_p = {\pi^2 \over 8} {1 \over m^2} (p m)^2\Gamma(pm)  \ee
by requiring that in the $m \rightarrow \infty$ limit keeping $z^2 = 8 m
r$ fixed (which is equivalent to looking at $r \ll m$), $H_p(r)$
becomes (using 13.3.3 of \cite{hmf})
\be H_p(z) = {\pi^2 \over 2 z^2}\,  p z K_1(pz)  \ee
whose Fourier transform is
\be \int {d^4 p \over (2 \pi)^4} e^{i p y}  H_p(z) = {1 \over (y^2 + z^2)^3}\ .  \label{smallr} \ee
We finally arrive at the statement
\be H(y,r) = 1 + Q_{M2} \int dp \, { (py)^2 J_1(py) \over 4 \pi^2 y^3}  H_p(r) \label{answer} \ee
where we have reduced (\ref{fourier}) to an integral over a single
variable by exploiting the spherical symmetry in $p$-space.

The solution (\ref{answer}) combined with the ansatz (\ref{ansatz}) is
the main result of this paper.  Dimensional reduction of this solution
along the $\psi$ coordinate of the Taub-NUT geometry (\ref{TNmetric})
will give rise to the solution type IIA supergravity describing D2
localized along the world volume of D6.  The metric part of the
solution is given by
\beq ds^2 &=& H(y,r)^{-1/2} \left( 1 + {2m \over r }\right)^{-1/2}
(-dt^2 + dx_1^2 + dx_2^2) \label{IIA}\\
&&
+ H(y,r)^{1/2} \left( 1 + {2m \over r}\right)^{-1/2} (dy^2 + y^2 d
\Omega_3^2)
+H(y,r)^{1/2}\left( 1 + {2m \over  r}\right)^{1/2}(dr^2 + r^2 d \Omega_2^2),\nonumber
\eeq
where $4m=R=g_s l_s$.
It is clear that when we set $Q_{M2}=0$, the solution reduces to the
supergravity solution containing only the D6-branes. Similarly, in the $m \rightarrow
0$ limit,
\be H(y,r) =  1 + {Q_{D2} \over (y^2 + r^2)^{5/2}}, \label{larger} \ee
where using $l_p=g_s^{1/3}l_s$,
\be Q_{D2} = {3 \over 64 m} Q_{M2} = 6\pi^2  g_s N_2 l_s^5, \ee
which agrees with the supergravity solution of the D2 by itself including the
numerical factors. Although (\ref{answer}) is left in an integral
form, the expression is completely explicit and the final integration
can be done numerically if desired. To demonstrate this point, we have
computed
\be f(r) = {512 m^3 r^3 \over Q_{M2} }(H(0,r)-1)\ee 
numerically. The normalization was chosen so that $f(r) =1$ for $r
\rightarrow 0$. The result of this computation is illustrated in
figure \ref{figa}.  The result clearly illustrates the cross-over
between asymptotics (\ref{smallr}) and (\ref{larger}) for small and
large $r$, respectively.

\FIGURE[t]{
\centerline{\rule{-6ex}{0ex}\raisebox{0.85in}{\rotatebox{90}{$\log(f(r))$} }
\includegraphics[width=3.5in]{potential.epsi}}
\centerline{$\log({r \over m})$}
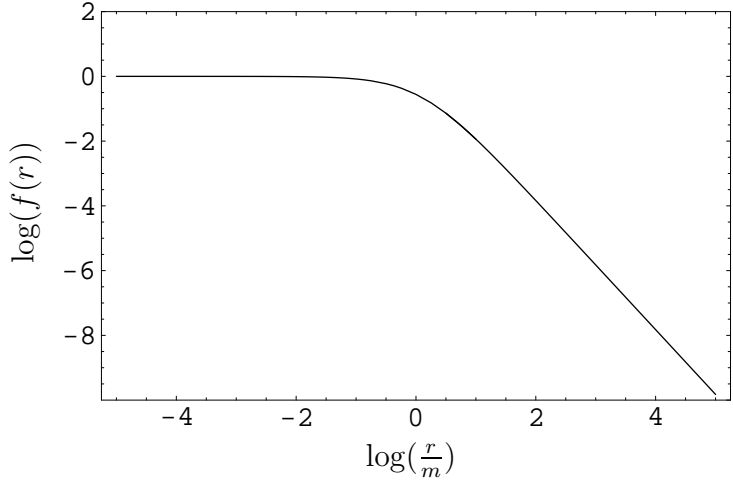
\caption{Log-Log plot of the function $f(r)$. \label{figa}}
}

\section{The Decoupling Limit \label{decouple}}

Now that we have constructed the supergravity solution of the localized
intersection of D2 and D6 in type IIA supergravity, let us consider
taking its near horizon limit which gives rise to a holographic dual
of the gauge theory on the D2-branes. We will scale $l_s$ to zero
keeping the two dimensional gauge coupling
\be g_{YM2}^2 = g_s l_s^{-1} \ee
fixed. In this limit, the gauge coupling on the six brane
\be g_{YM6}^2 = (2 \pi)^4 g_s l_s^3 = (2 \pi l_s)^4 g_{YM2}^2 \ee
goes to zero so that the dynamics on the six-brane decouples.  In order
to identify the corresponding near horizon geometry on the
supergravity side, we also scale the radial coordinates so that
\be Y = {y \over l_s^2}, \qquad U = {r \over l_s^2} \ee
is fixed.  In this limit, the harmonic function due to the  D6-brane source
scales as
\be 1 + {2 m \over r} = 1 + {g_s l_s  \over 2 r} = 1 + {g_{YM2}^2  \over 2 U} \ ,  \ee
whereas the harmonic function due to the D2-brane source
(\ref{answer}) scales as
\be H(Y,U) = {1 \over l_s^4} h(Y,U) \ , \ee
where
\beq
\lefteqn{ h(Y,U) =}\\
&& \pi^2 g_{YM2}^4 N_2 
\int dP 
{(PY)^2 J_1(PY) \over  Y^3} P^2 \Gamma \left({g_{YM2}^2  P \over 4} \right)
e^{-PU} \U(1+{g_{YM2}^2 P \over 4},2,2PU) \ .  \nonumber 
\eeq
We have also scaled the integration variable
\be p = {P \over l_s^2}  \ee
so that the string length $l_s$ does not appear anywhere in the
definition of $h(Y,U)$. 

We now have all the ingredients to explicitly write down the
supergravity solution for the decoupled D2/D6 system. Generalizing
slightly to the case with multiple coincident D6-branes, the solution
takes the form\\
\parbox{\hsize}{
\beq 
\lefteqn{{ds^2 \over l_s^2}  = h(Y,U)^{-1/2} \left(1 + {g_{YM2}^2 N_6\over 2U}\right)^{-1/2}
(-dt^2 + dx_1^2 + dx_2^2)} \nonumber \\
&&
+ h(Y,U)^{1/2} \left(1 + {g_{YM2}^2 N_6\over 2U}\right)^{-1/2} (dY^2 + Y^2 d
\Omega_3^2) \label{sugra} \\
&& +h(Y,U)^{1/2}\left(1 + {g_{YM2}^2 N_6\over 2U}\right)^{1/2}(dU^2 + U^2 d \Omega_2^2)\nonumber  \\
\lefteqn{h(Y,U) =}  \\
&&\pi^2 g_{YM2}^4 N_2 
\int dP 
{(PY)^2 J_1(PY) \over  Y^3} P^2 \Gamma \left({g_{YM2}^2 N_6 P \over 4} \right)
e^{-PU} \U(1+{g_{YM2}^2 N_6 P \over 4},2,2PU) \ . \nonumber 
\eeq
}

The only dependence of this metric on the string length $l_s$ is in
the overall normalization which is what one expects for the
supergravity dual of a quantum field theory. Note also that although
we no longer have the ``1'' in the harmonic function of the D2-brane in
taking the decoupling limit, we are left with the ``1'' in the harmonic
function of the D6-brane.  The effect of the D2-brane is therefore to
``warp'' not only the ALE region but also the asymptotically flat
region of the D6-brane geometry.

\section{RG Flow and Mirror Symmetry}

Let us now interpret the various features of the supergravity solution
(\ref{sugra}) from the point of view of the field theory dual. By
construction, (\ref{sugra}) is dual to maximally supersymmetric SYM in
2+1 dimensions with gauge group $SU(N_2)$ further coupled to $k=N_6$
flavors of massless hypermultiplets in the fundamental representation.
The coupling to the hypermultiplets reduces the number of unbroken
supersymmetries from 16 to 8. The supergravity solution (\ref{sugra})
asymptotes to the geometry of the near horizon D2-brane in the large
$U$ limit.  This suggests that the dynamics of this theory is
dominated by the free gluons in the UV as one expects for a
superrenormalizable theory.  In the small $U$ region, the geometry of
(\ref{sugra}) asymptotes to $AdS_4 \times S_7 / \Z_{k}$. This is the
superconformal field theory one expects to find on M2-brane probing
$\R^4 \times (\R^4 / \Z_{k})$.

Now, the superconformal theory on the M2-brane on an orbifold does not
have a simple Lagrangian formulation. One way to define such a theory
without relying on string theory is to define it as an IR fixed point
of a different theory which has a Lagrangian formulation.  The 2+1 SYM
with $k$ flavors is one concrete example of a UV theory which flows to
this superconformal field theory in the IR.  Roughly speaking, by
compactifying one of the directions of $\R^4 / \Z_{k}$ into a Taub-NUT
space, one has embedded the dynamics of M2 into the dynamics of D2,
and it is the latter which has a good Lagrangian description.  By
making the D2 warp the $\R^4 \times \mbox{Taub-NUT}$ geometry due to
its gravitational backreaction, one constructs the supergravity dual to
the decoupled theory on the D2-brane. The supergravity solution
(\ref{sugra}) encodes this full renormalization group flow in the
language of AdS/CFT correspondence. Similar observations can be found
in the earlier work of \cite{Pelc:1999ms}.

It turns out that there is a different way to embed the superconformal
field theory on the M2 on $\R^4 \times (\R^4 / \Z_{k})$ as the IR
fixed point of a field theory with a Lagrangian description. This is the
$\Z_k$ quiver theory of the 2+1 SYM. From the point of view of branes,
this amounts to considering the decoupling limit of M2-branes on
$\R^3\times S_1 \times (\R^4 / \Z_k)$.  The supergravity solution for
such a brane configuration is easy to find. They are simply
\be ds^2 = H^{2/3} (-dt^2 + dx_1^2 + dx_2^2) + H^{1/3}(dy^2 + y^2 d \Omega_2^2 + dz^2 + dr^2 + r^2 ds_{Lens}^2) \ee
where $ds^2_{Lens}$ is the metric on the Lens space which is the base
of the $\R^4 / \Z_k$ viewed as a cone, and
\be H(y,z,r) =  1 + \sum_{n=-\infty}^\infty {Q_{M2}  \over (r^2+y^2 + (z - 2 \pi n R )^2)^3} \ . \ee
It is possible to take the decoupling limit of this solution keeping
\be {R \over l_s^2}  = {g_s \over l_s} = g_{YM2}^2 = \mbox{fixed} \ee
which will give rise to a different supergravity background describing
the renormalization group flow of the 2+1 dimensional $\Z_k$ quiver
theory flowing to the same superconformal field theory.

What we have here is a pair of supergravity solutions, both of which
asymptotes to the same $AdS_4 \times S_7 / \Z_k$.  It is therefore a
holographic realization of two different RG flows which flow to the
same conformal field theory in the far IR. This is mirror
symmetry. Although the metric on the supergravity side asymptotes to
the same thing near the core, the geometry away from the core of the
two solutions are clearly different from each other. This illustrates
quite explicitly in the AdS/CFT language the basic fact that mirror
symmetry is an equivalence only for the far IR of a pair of field
theories.

The basic idea behind the embedding of the superconformal field theory
into a Lagrangian field theory was to compactify one of the
dimensions either in the $\R^4$ or the $\R^4/\Z^k$. The freedom to
choose between the two was the basis for mirror symmetry.  Let us now
consider what happens if one compactifies both so that we have $(\R^3
\times S^1) \times \mbox{Taub-NUT}$. Now there are two ways to reduce
the same geometry from M-theory to type IIA.  Let us for the sake of
the argument reduce on the circle in the Taub-NUT. This will give rise to
a D2/D6 system with one of the direction transverse to the D2-brane
compactified.  In the decoupling limit, compactness of the directions
transverse to the world volume of the brane is an indication that the
underlying gauge theory is a $U(\infty)/\Z$ theory because of the
presence of images. According to the argument of \cite{Taylor:1997ik}, it is
better to view this as a theory with one extra dimension.  From the
point of view of the supergravity, the same picture manifests itself
in the fact that in the near horizon limit, where the backreaction of
the D2-brane dominates, the proper size of the $S^1$ transverse to the
D2 shrinks as one approaches the boundary. At the point where this
proper size becomes smaller than the string length, the supergravity
description of this geometry become unreliable, and following the
argument of \cite{Itzhaki:1998dd}, one is instructed to go to the
T-dual picture, where the D2 becomes a D3.

Unfortunately, the same T-duality maps the D6 to a D5. T-duality in
supergravity is not capable of handling this map except for the case
where the D5 is completely smeared.  We are therefore unable to
provide a purely supergravity description of the 3+1 dimensional UV
fixed point for the decoupled theory on M2 in $(\R^3 \times S^1)\times
\mbox{Taub-NUT}$. The problem of finding this supergravity solution
was attempted most recently in \cite{Fayyazuddin:2002bm}. Let us note
in passing that at least the solution for the case of the smeared D5
considered in section 4.1 of \cite{Fayyazuddin:2002bm} can be obtained
from (\ref{IIA}) by applying the T-duality rules for supergravity.

Another interesting aspect of the decoupled theory of M2 on
$(\R^3\times S^1) \times \mbox{Taub-NUT}$ is the fact that its UV
description on the field theory side is precisely the defect field
theory introduced in \cite{Dasgupta:1999wx,Karch:2000gx}. In fact, one
of the main motivations of \cite{Fayyazuddin:2002bm} was to find the
purely gravitational holographic description of the defect field
theory.  By reducing from M-theory to IIA on the circle of the
Taub-NUT and T-dualizing to IIB on the circle in $\R^3 \times S^1$, we
arrive at a defect theory consisting only of the D5 defects.  The
Lagrangian of this theory was worked out in
\cite{DeWolfe:2001pq}. Alternatively, one could have reduced from
M-theory to IIA on the circle in $\R^3 \times S^1$ and T-dualizing along
the circle of the Taub-NUT arriving at the theory with NS defects.
The relation between the two ways of going from M-theory to type IIB
is the natural extension of mirror symmetry.  Clearly, from the point
of view of the type IIB theory, this equivalence is S-duality.  What
one learns here is that the origin of mirror symmetry in 2+1
dimensional gauge theory is the S-duality of the defect field theory
in 3+1 dimensions.

This idea of embedding a 2+1 dimensional theory in some UV structure
to make mirror symmetry manifest is not a new idea.  Embedding to
string theory was exploited for this goal some time ago in
\cite{Porrati:1997xi,Hanany:1997ie}.   Embedding of the 2+1
dimensional theory into the 3+1 dimensional defect field theory amounts to
taking the decoupling limit of \cite{Hanany:1997ie}.  Although the
formulation of defect field theories was strongly motivated by string
theory, their existence is independent of string theory. The relation
between S-duality of the defect field theory and the mirror symmetry
of its dimensionally reduced theory can be studied by exploiting their
holographic duality, instead of their embedding, to string theory. The
former is physically economical.

By embedding a pair of 2+1 dimensional mirror theories into a pair of
S-dual defect field theories, one obtains a mirror duality which applies at
all scales, not just in the far IR.  So the embedding into the defect
field theories can be interpreted as an intricate UV modification of
the mirror pair theories so as to extend their range of validity
beyond the far IR.  For the Abelian case, this issue was addressed in
\cite{Kapustin:1999ha}. For the non-Abelian case, the natural UV
modification appears to involve a theory with one extra dimension and
defects.

\section{Some Generalizations}

In the previous sections, we discussed mainly the localized D2/D6
supergravity configuration where all of the D2 and D6 are
coincident. Let us now consider some generalizations.

\subsection{Separating D2 from D6}

One simple generalization one can consider is separating the D2-brane
from the D6-brane. For the sake of concreteness, let us consider the
case where there is one of each of D2 and D6.

{}From the point of view of M-theory, this is a configuration of a
single M2-brane at a generic point on the Taub-NUT background
(\ref{TNmetric}). Therefore, the supergravity solution is given by the
same ansatz (\ref{ansatz}) where $H$ is a solution of the harmonic
equation (\ref{harm}) but with a source located at a generic point in
the Taub-NUT.  From the point of view of finding a localized D2/D6
solution of the type IIA supergravity equation of motion, however, one
is only interested in sources which are smeared along the 11-th
coordinate $\psi$.  Let us therefore take the M2-brane source to be
smeared evenly along $\psi$ as this would also simpilfy the analysis.
The harmonic function $H$ is then a solution of
\be (\nabla^2_{TN} + \nabla^2_y) H =
2 \pi^4 Q_{M2}  \delta^4(\vec y - \vec y_0) \delta^3(\vec r - \vec r_0) \delta(\psi - \psi_0) \ . \ee
The factor of $2\pi^4$ arises from the fact that
\be \nabla_r^2 \left( {1 \over r^6} \right) = 2 \pi^4  \delta^8(r) \ee
in 8 dimensions. Smearing along $\psi$ and separating the variables as
was done in (\ref{fourier}), one finds that $H_p(\vec r)$ is a
solution of
\be \left(\nabla_r^2  - {2 m p^2 \over |\vec r |} - p^2\right)
 H_p(\vec r) = {2 \pi^4 Q_{M2}  \over 2 \pi R} \delta^3(\vec r - \vec r_0) ,
\qquad R = 4m \ee
with appropriate boundary conditions. This equation is  of the form
\be \left(\nabla_r^2 + {2 k \nu \over |\vec r|} + k^2 \right) G(\vec r, \vec r_0) = \delta^3 (\vec r - \vec r_0) \ee
if one identifies
\be H_p(\vec r) = {2 \pi^4 \over 2 \pi R} G(\vec r, \vec r_0), \qquad \nu = i m p, \qquad k = i p \ . \label{identification} \ee
Precisely this equation with the appropriate boundary conditions was
considered in \cite{Hostler} and the solution was found to be
\be G(\vec r, \vec r_0) =
 - {\Gamma(1 - i \nu) \over 4 \pi |\vec r - \vec r_0|} 
{1 \over ik}
\left( -{\partial \over \partial y} + {\partial \over \partial x}\right)
\W_{i \nu, 1/2}(-i k x) \M_{i \nu, 1/2}(-i k y) \label{green} \ee
where $\M_{a,b}(z)$ and $\W_{a,b}(z)$ are the Whittaker functions, and 
\be x = r + r_0 + | \vec r - \vec r_0|, \qquad  y = r + r_0 - | \vec r - \vec r_0| \ . \ee
If one sends the source $\vec r_0$ to zero, $G(\vec r, \vec r_0)$ simplifies to
\be G(\vec r) = {1 \over 4 \pi r} \Gamma(1 - i \nu) \W_{i \nu, 1/2} (- 2 i k r) \ . \label{sourceorigin} \ee
Using the identity
\be 
\W_{\kappa,\mu}  = z^{\mu+1/2}  e^{-z/2}  \U(\mbox{${1 \over 2}$} + \mu - \kappa, 1 + 2 \mu, z) \ee
and the identification (\ref{identification}), one can show that
(\ref{sourceorigin}) is equivalent to (\ref{analyticform}) including
all the numerical factors.

Using (\ref{green}) and the identification (\ref{identification}), one
can write an explicit expression for the supergravity solution of the
localized D2/D6 configuration. Since harmonic equations are linear, it
is straightforward to generalize this to the case where the D2 is
distributed arbitrarily in transverse coordinates.  It is also
straightforward to generalize this solution to the case where there
are multiple D6's as long as all of the D6's are coincident. Simply set
\be \nu = i N_6 m p  \ . \ee

By scaling $\vec r_0= \alpha' \vec U_0$ keeping $\vec U_0$ fixed, one
can take the decoupling limit of the D2/D6 as we did in section
\ref{decouple}.  From the point of view of the field theory dual, this
corresponds to turning on a vacuum expectation value of some of the
adjoint scalars, so that the matter fields in the fundamental acquire
 mass.

\subsection{Separating the D6}

Another possible generalization one might consider is to separate the D6's from one another. The M-theory lift of this configuration is the multi-centered Taub-NUT geometry whose metric is given by
\be ds^2 = V d \vec r^2 + {(4m)^2 \over V}(d \psi + \vec{\omega} \cdot d \vec r)^2 \ee
where
\be V(\vec r) = 1 + \sum_{i = 1}^{N_6} {2 m \over |\vec r - \vec r_i|}, \qquad \nabla V = 4m \nabla \times \vec \omega \ . \ee

Applying the same ansatz as (\ref{ansatz}) gives rise to a new
harmonic equation (\ref{harm}).  Although it is not absolutely
necessary to do so, let us smear the M2 source along the $\psi$
coordinates. This will simplify the analysis. Then,
the harmonic equation (\ref{harm}) can be written explicitly 
as
\be \left( \nabla_r^2 - p^2 V(\vec r)\right) H_p(\vec r) = {2 \pi^4 Q_{M2} \over 2 \pi R} \delta^3(\vec r - \vec r_0) \ . \label{harm2} \ee

For a geneneral multi-centered Taub-NUT background, this is still a
difficult equation to solve.  The case of all the D6 being 
coincident gives rise to a single centered Taub-NUT considered in the previous section.

It turns out that double centered Taub-NUT also admits natural
coordinates in which the harmonic equation (\ref{harm2}) separates  \cite{Gibbons:1988sp}. 
Let us consider this case in some detail.

Consider a double centered Taub-NUT with $N_1$ coincident centers at
$\vec r_1$ and $N_2$ coincident centers at $\vec r_2$. Without loss of
generality, we can set $\vec r_2 = - \vec r_1$.  One can then
introduce the so called prolate spheroidal coordinates
\be \xi = {|\vec r - \vec r_1|  + |\vec r - \vec r_2| \over 2L}, \qquad
\eta = {|\vec r - \vec r_1|  - |\vec r - \vec r_2| \over 2L}
\ee
where $2L = |\vec r_1 - \vec r_2|$ is the distance between the two
centers. The ranges of these coordinates are $1 < \xi$ and $-1 < \eta <
1$.  The countours of fixed $\xi$ and $\eta$ are illusterated in
figure \ref{figb}.

\FIGURE[t]{
\centerline{\includegraphics[width=3.5in]{ellipsoidal.epsi}}
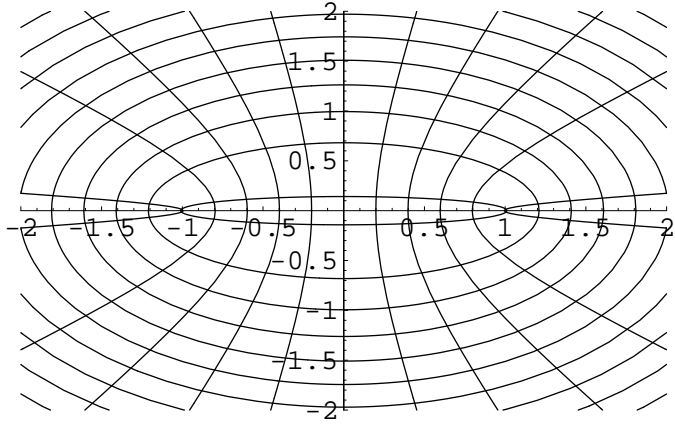
\caption{Contours of fixed prolate spheroidal coordinates $\xi$ and $\eta$. \label{figb}}
}

Let $\phi$ denote the angular coordinate around a symmetry axis
defined by $\vec r_1 - \vec r_2$.  Then, the set of coordinates $\xi$,
$\eta$, and $\phi$ specifies a point $\vec r$. In these coordinates,
the harmonic equation (\ref{harm2}) becomes
\beq&&\left( \partial_\xi (\xi^2 - 1) \partial_\xi + 
\partial_\eta (1 - \eta^2) \partial_\eta  +
\left({1 \over \xi^2 - 1} + {1 \over 1 - \eta^2}\right) \partial_\phi^2  - L^2 (\xi^2 - \eta^2) V p^2\right) H_p(\xi,\eta,\phi)  \nonumber \\
&&  \qquad =  {2 \pi^4 Q_{M2} \over 2 \pi L R} \delta(\xi - \xi_0) \delta(\eta - \eta_0) \delta(\phi - \phi_0) \eeq
where
\be (\xi^2 - \eta^2) V = (\xi^2 - \eta^2) + {2 m (N_1+ N_2) \xi  \over L}  - {2 m (N_1- N_2) \eta  \over L}  \ . \ee

To further analyze this problem, it is convenient to consider the solution 
to the equation
\be\label{prop} \left(-\partial_\eta (1 - \eta^2) \partial_\eta 
+{k^2 \over 1 - \eta^2} 
- p^2 L^2 \eta^2 -  2 p^2 m L (N_1 - N_2) \eta  \right)
B_{\lambda, k}(\eta) = \lambda B_{\lambda, k}(\eta) \ . \ee
This equation takes the form
\be \label{acsf}
\frac{1}{\sin\theta}\frac{d}{d\theta}\sin\theta\frac{d}{d\theta}B_{\lambda,k}+\left(\lambda-\frac{k^2}{\sin^2\theta}+
2p^2mL(N_1-N_2)\cos\theta+p^2L^2\cos^2\theta\right)B_{\lambda,k}=0 \ee
after making the change of variables
\be \eta = \cos \theta , \qquad 0 < \theta < \pi. \ee
In the $L \rightarrow 0$ limit, this equation becomes the Legendre
equation.  For finite $L$, one expects a discrete spectrum of
eigenvalues $\lambda_n$ and its associated eigenfunction
$B_{\lambda_n,k}(\eta)$.  They can be determined either using
numerical methods, perturbation theory, or by expanding
\be B_{\lambda, k}(\eta)=e^{pL\eta}\sum_{s=0}^\infty c_s P_{s+k}^k(\eta),
\ee
and deriving a recursion relation for the coefficients $c_s$ along the
lines of \cite{BH}. We will consider these eigenfunctions to be
orthonormalized so that
\be \int_0^\pi d\theta \sin(\theta) \, B_{\lambda_m, k}(\cos\theta) B_{\lambda_{n}, k}(\cos\theta) = 
\int_{-1}^1 d \eta \, B_{\lambda_m, k}(\eta) B_{\lambda_n, k}(\eta)  
= \delta_{mn}. \ee

One now sees that upon parameterizing
\be H_p(\xi,\eta,\psi) = \sum_{n,k} A_{\lambda_n,k}(\xi) B_{\lambda_n,k}(\eta) e^{ik\phi}\ , \ee
where $B_{\lambda_n,k}$ are orthonormal, $A_{\lambda_n,k}(\xi)$ satisfies
\beq && \left(\partial_\xi (\xi^2 - 1) \partial_\xi -{k^2 \over \xi^2 - 1}  - p^2 (L^2\xi^2 + 2 m (N_1+N_2) L \xi ) - \lambda_n \right) A_{\lambda_n,k} (\xi) \nonumber \\
&& \qquad = {2 \pi^4 Q_{M2} \over 2 \pi L R} B_{\lambda_n,k}(\eta_0) e^{-k \phi_0} \delta(\xi - \xi_0) \ . \eeq
Although somewhat complicated, this is a linear inhomogenious ordinary
differential equation which can be solved  numerically or
using the method of  \cite{BH}. One can then evaluate
\be H(y,\xi,\eta,\phi)=  1 + \int {d^4 p \over (2 \pi)^4}
e^{i  p y}\, 
\sum_{n,k} A_{\lambda_n,k} (\xi) B_{\lambda_n,k}(\eta) e^{ik \phi} \ee
which when substituted into (\ref{ansatz}) gives rise to the
supergravity solution of the D2 parallel to two collections of D6-branes.

\section{Concluding Remarks}

The main goal of this paper was the construction of the localized
D2/D6 supergravity solution.  We have constructed this solution
explicitly as an integral expression in (\ref{answer}). Using this
solution, it was possible to construct a holographic dual to 2+1
dimensional Yang-Mill with matter in the fundamental representation,
and describe mirror symmetry in the language of AdS/CFT
correspondence.

It would be interesting to explore the standard holographic
observables: entropy, Wilson loop, and correlation functions, for this
supergravity solution.  It would also be interesting to explore the
mapping of observables between the mirror pairs from the point of view of
holography.

The D2/D6 system appears to be unique in providing a conceptually
clean setup to add flavors to AdS/CFT. The D3/D7 system suffers from
the lack of asymptotic freedom, and the D1/D5 system suffers from the
lack of moduli-space. One can still describe the decoupled theory on
D1/D5 in Born-Oppenheimer approximation, but that does not appear to
be compatible with the holographic duality, which has as a starting point a
stationary supergravity solution with definite configuration of static
sources.  One manifestation of this difficulty is the fact that a
localized supergravity solution of D1 coincident with D5 does not even
appear to exist \cite{Marolf:1999uq}.

The key to the simplicity is the fact that the D6-brane lifts to a
Taub-NUT geometry in M-theory which is purely geometrical.
Furthermore, the geometry is sufficiently regular both near the core
and at infinity.  This is what made generalization of the D2/D6
intersection in the near core region considered in \cite{Itzhaki:1998uz} to
the full Taub-NUT geometry possible. It would be very interesting to
see if the D4/D6 intersection considered in \cite{Hashimoto:1998ug}
can be extended in a similar manner by taking advantage of the
simplicity of the Taub-NUT geometry.

A different simple generalzation is the supergravity dual of the
non-commutative gauge theory with fundamental matter.  Such a solution
can be found by applying the same T-duality transformation considered
in \cite{Hashimoto:1999ut} or by following the twist operation for the
dipole theories introduced in \cite{Bergman:2001rw}. The resulting
supergravity background is similar to (\ref{sugra}) but with the
metric along the D2-brane worldvolume replaced by
\be {ds^2 \over l_s^2}  = h(Y,U)^{-1/2} \left(1 + {g_{YM2}^2 N_6\over 2U}\right)^{-1/2}
\left(-dt^2 + {dx_1^2 + dx_2^2 \over 1 + \Delta^4 h(Y,U)^{-1} \left(1 + {g_{YM2}^2 N_6\over 2U}\right)^{-1}}\right) + \ldots\ee
where $2 \pi \Delta^2 = \theta^{12}$ is the non-commutativity parameter along
the D2-brane world volume.

There are other generalizations one might consider.  For example, one
can separate the D6's completely, so that the M-theory background
becomes that of a multi-center Taub-NUT. One might also consider
finding the supergravity solution for a decoupled theory on M2 probing
a manifold of $Sp(2)$ holonomy\footnote{The theory on the brane probe
was considered in \cite{Gukov:2002es}.  Mirror symmetry for these
theories were first considered from the point of view of brane
constructions in \cite{Kitao:1998mf,Lee:1999ze,Kitao:1999uj}.}
considered in \cite{Gauntlett:1997pk}. For these cases, separation of
variables do not appear to work and more sophisticated methods for
solving for the the harmonic function must to be employed.

\section*{Acknowledgements}
We would like to thank
K.~Hori,
N.~Itzhaki,
A.~Kapustin,
A.~Karch, 
D.~Marolf, 
and
S.~Mukhi
for discussions.  We also thank the Aspen Center for Physics and 
Institute des Hautes \'Etudes Scientifiques where
part of this work was done.  This work is supported in part by DOE
grant DE-FG02-90ER40542 and by the  Marvin L.~Goldberger fellowship.

\bibliography{taubnut}\bibliographystyle{utphys} \end{document}